 \documentstyle[prl,aps,epsfig,preprint]{revtex}

\newcommand{\dd}{\partial}

\newlength{\textwidthm}
\begin{document}

\draft

 \preprint{Group Delay-NLO}

\title{
	Ultra-Slow Light and Enhanced Nonlinear Optical
	Effects in a Coherently Driven Hot Atomic Gas.
}
\author{
	Michael~M.~Kash,$^{1,5}$
	Vladimir~A.~Sautenkov,$^{1}$
	Alexander~S.~Zibrov,$^{1,3}$
	L.~Hollberg,$^{3}$
	George~R.~Welch,$^{1}$
	Mikhail~D.~Lukin,$^{4}$
	Yuri~Rostovtsev,$^{1}$
	Edward~S.~Fry,$^{1,2}$
	Marlan~O.~Scully$^{1,2}$
}
\address{
	$^1$ Department of Physics, Texas A\&M University,
		College Station, Texas~~77843-4242 \\
	$^2$ Max-Planck-Institut f\"{u}r Quantenoptik,
		D-85748 Garching, Germany \\
	$^3$ National Institute for Standards and Technology,
		Boulder, Colorado~~80303 \\
	$^4$ ITAMP, Harvard-Smithsonian Center for Astrophysics,
		Cambridge, Massachusetts~~02138 \\
	$^5$ Department of Physics, Lake Forest College,
		Lake Forest, Illinois~~60045
}
\date{\today}
\maketitle

\begin{abstract}

	We report the observation of small group velocities
of order 90 meters per second, and large group delays of
greater than 0.26 ms, in an optically dense hot rubidium gas
($\approx 360$ K).  Media of this kind yield strong nonlinear
interactions between very weak optical fields, and very sharp
spectral features. The result is in agreement with previous
studies on nonlinear spectroscopy of dense coherent media.

\end{abstract}


 \newpage

	A phase coherent ensemble of atoms (``phaseonium,'')
represents a truly novel state of matter.  A dramatic
example of such a quantum coherence effect is provided by
the recent report of extremely slow group velocity (17 m/s)
for a pulse of light in a Bose condensate of ultra-cold
sodium atoms~\cite{hau99nature}.  Previous direct and indirect
measurements of increasingly low group velocities in coherently
prepared media have ranged from c/165~\cite{Harris-slow}, and
c/3000~\cite{wynands} to about c/$10^6$~\cite{lukin97prl}.
In this report we show that by a proper choice of experimental
parameters such as atomic density and optical intensity,
very large group delay (slow group velocity) of light can be
observed in a cell of hot (360 K) $^{87}$Rb atoms.  This is
in agreement with our previous studies of spectroscopy and
magnetometry in dense coherent media.


	Furthermore, we demonstrate here that this
relatively easily created medium also displays very
strong nonlinear coupling between very weak optical
fields~\cite{Harris-NLO-weak}.  Specifically, with such a
thermal ensemble of rubidium atoms, we observe (1) group
delay ($T_g$) of 0.26 ms for propagation through our 2.5
cm long, optically thick, electromagnetically-induced
transparent (EIT) medium, and (2) extremely efficient nonlinear
interactions.  These two aspects of phaseonium are closely
related: $T_g$ will be shown to be the figure of merit for
various linear and nonlinear optical processes using EIT (see
e.g. Eq.~(\ref{why_this_one}) and Ref.\cite{Harris-NLO-weak}).
Specific manifestations of these unusual properties of
dense coherent media include  new regimes of high precision
spectroscopy and nonlinear interactions of the very weak light
fields ~\cite{harris-yam-98-prl,nonlinear-lowlight} with greatly
alleviated phase matching requirements~\cite{phasematch}.

\begin{figure}[ht]
 \centerline{\epsfig{file=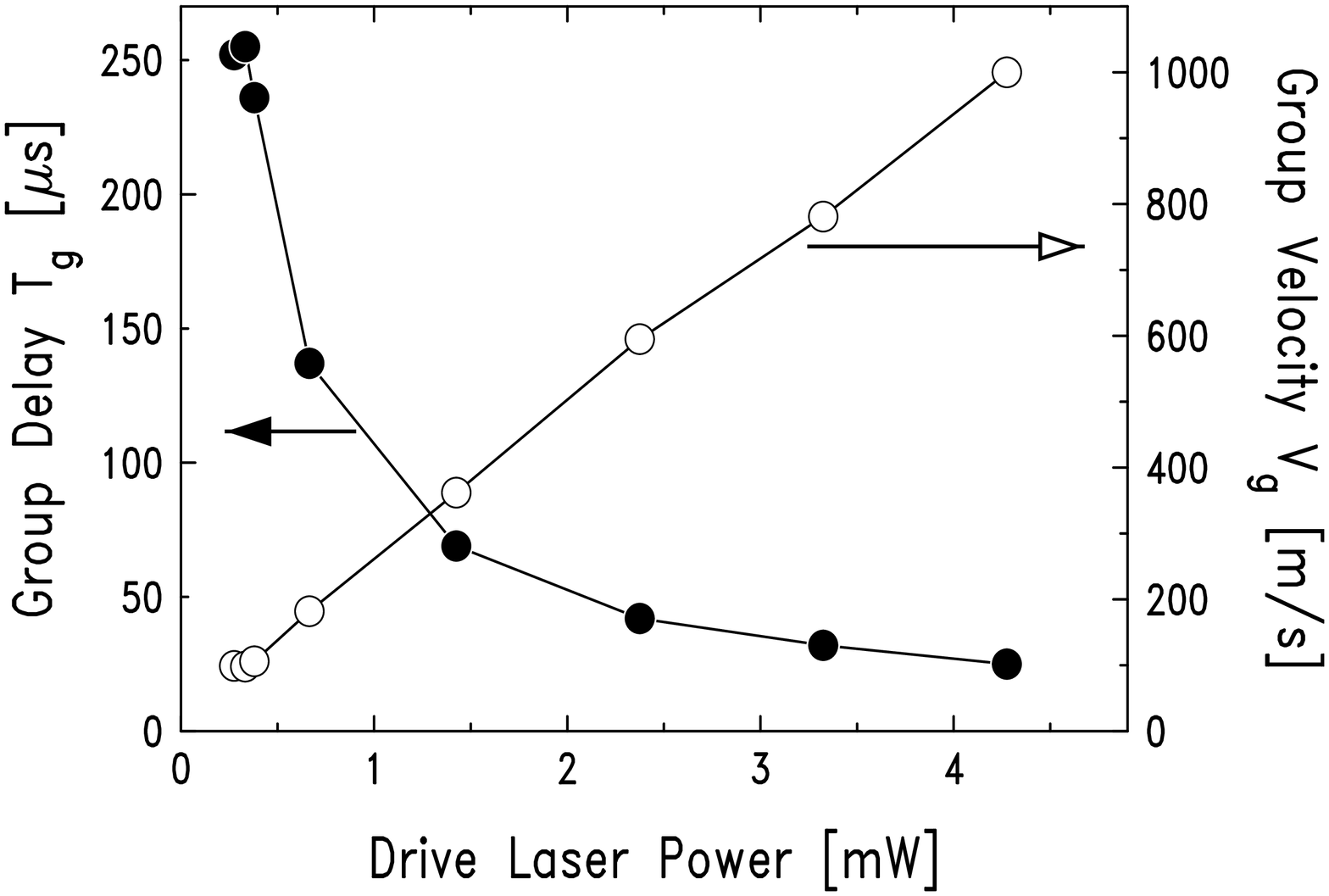,width=8.6cm}}
 \vspace*{2ex}
 \caption{
 	\label{grp-data.fig}
	Observed group delay (solid circles) and average
	group velocity (open circles) as a function of
	the drive laser power.	The density of $^{87}$Rb
	was $2\times 10^{12}~{\mathrm cm}^{-3}$ and the
	laser beam diameter was 2 mm.  For this transition,
	$\Omega/(2\pi) = 1\times 10^6 \sqrt{I}$ where $I$
	is in mW/$\mathrm{cm}^2$.
 }
\end{figure}

\begin{figure}[ht]
 \centerline{\epsfig{file=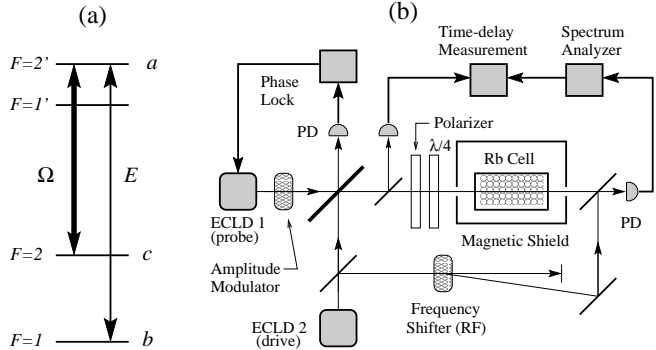,width=8.6cm}}
 \vspace*{2ex}
 \caption{
	\label{levs-expt.fig}
	(a) Level scheme for group delay measurement.
	(b) Schematic of the experiment.  The PDs represent
	high speed photo-detectors.  Amplitude Modulator and
	Frequency Shifter are acousto-optic modulators.
 }
\end{figure}

	We observed large group delay on the D1 resonance
line ($\lambda$=795 nm) of $^{87}$Rb (nuclear spin I=3/2).
The cell contained isotopically pure $^{87}$Rb and 30 Torr
of Ne buffer gas.  Under this condition, with a 2 mm laser
beam diameter, the ground-state coherence relaxation rate
$\gamma_{bc}/(2\pi)$ is reduced below 1 kHz.  The measured time
delay as a function of the power of the drive input to the cell
is shown in Fig.~\ref{grp-data.fig}.  The drive laser was tuned
to the 5$^2$S$_{1/2}$(F=2)$\rightarrow$5$^2$P$_{1/2}$(F=2)
transition; a co-propagating probe laser was tuned to the
5$^2$S$_{1/2}$(F=1)$\rightarrow$5$^2$P$_{1/2}$(F=2) transition
(see Fig.~\ref{levs-expt.fig}a).  Both of these lasers were
external cavity diode lasers.  They were phase-locked with
a frequency offset near the ground-state hyperfine splitting
of 6.8 GHz, which was fixed by a tunable microwave frequency
synthesizer.  The probe laser power was 5\% of the drive
laser power, and was amplitude modulated by approximately 50\%
with a sine wave at a frequency that was varied in the range
of 0.1--10~kHz.

\begin{figure}[ht]
 \centerline{\epsfig{file=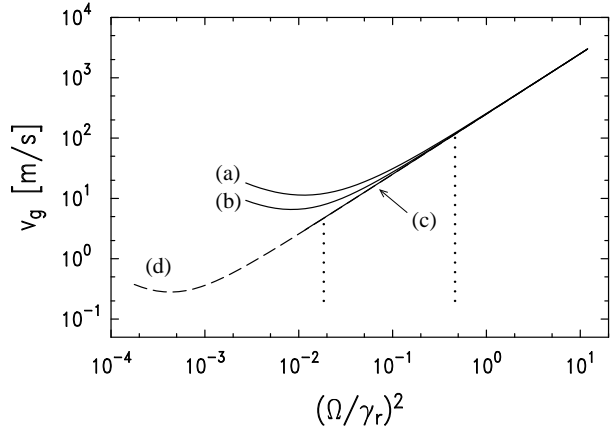,width=8.0cm}}
 \vspace*{2ex}
 \caption{
 	\label{dopp.fig}
	Calculated group velocity versus drive laser power for
	Rb with density $N = 2\times 10^{12}~\mathrm{cm}^{-3}$.
	Curves (a), (b), and (c) are calculated with lorentzian,
	gaussian, and no Doppler averaging respectively.
	A ground state relaxation rate $\gamma_{bc}/(2\pi) =
	1000$~Hz, is assumed for all three.  A collisionally
	broadened homogeneous half-width $\gamma/(2\pi) =
	150$~MHz and Doppler half-width $\Delta\omega_D/(2\pi)
	= 270$~MHz is used for curves (a) and (b), and the
	radiative half-width $\gamma/(2\pi) = \gamma_r/(2\pi) = 3$~MHz is used for
	curve (c).  For drive intensities below the
	right
	vertical dotted line, the medium is strongly opaque
	(absorption $>$ 90\%).	A key point is that for
	transparent media, the effect of Doppler averaging on
	$v_g$ is small.
	The dashed curve (d) is for the hot case as in (b),
	but with $\gamma_{bc}/(2\pi) = 40$~Hz, which is
	experimentally possible.  For such a lower coherence
	decay, the onset of strong absorption should occur at
	lower intensity,
	indicated by the left vertical dotted line,
	allowing for much lower $v_g$.
}
\end{figure}

	Figure \ref{grp-data.fig} also shows the inferred
{\it average} group velocity for each measured delay time.
As the lasers propagate down the length of the cell, the
drive laser power is attenuated.  Since the group velocity
decreases with drive laser power (see Eq.~(\ref{ng-on-omega})
and Fig.~\ref{dopp.fig}), we see that the instantaneous velocity
is lower toward the output end of the cell than near the input.
Hence, we report the average velocity in the cell.

	The group delay in passing through the cell was
measured by observing the time retardation of the amplitude
modulation upon passing through the cell.  The attenuation and
time delay was measured for a range of modulation frequencies,
allowing us to model the propagation for a wide range of pulses.
The time delay was independent of the modulation frequency
up to the linewidth of the EIT resonance.  Systematic effects
resulting in unwanted phase shifts of the amplitude modulation
of the light were investigated by several approaches: cooling
the cell so that very little Rb vapor was present, tuning far
from resonance, removing the cell, and checking the electronics
for spurious phase shifts.  A schematic of the experimental
setup is shown in Fig.~\ref{levs-expt.fig}b.

	In the experiment, both the drive beam and the probe
beam are transmitted through the cell, and because of nonlinear
optical processes additional frequencies are generated by the
medium as discussed below.  To isolate the amplitude of the
transmitted probe, we split off part of the drive before the
cell and shift its frequency down by a small amount (50 MHz)
as indicated in Fig~\ref{levs-expt.fig}b.  This shifted beam
bypasses the cell and is combined on the detector along with
the transmitted drive and probe and any generated fields.
Because the amplitude of the shifted field is constant, this
signal is proportional to the transmitted probe without any
contribution from the transmitted drive field.

	The low group velocity arises from the large
dispersion of the coherent medium.  For a light field with
a slowly changing complex amplitude, we write $E(z,t) =
{\cal E}(z,t) \exp(i kz - i\nu t)$, and $P(z,t) = {\cal
P}(z,t) \exp(i kz - i\nu t)$, where ${\cal E}(z,t)$ and
${\cal P}(z,t)$ are the slowly varying envelopes of the
electric field and atomic polarization.  The carrier wave has
wavenumber $k$ and frequency $\nu$.  The Fourier components
of the field and the polarization are related by $P(z,\nu)
= \epsilon_0\chi(\nu)E(z,\nu)$, where $\chi(\nu)$ is the
susceptibility of the medium.  Substituting these relations
into the wave equation and neglecting all derivatives greater
than the first, we obtain the equation of motion for the envelope:
\begin{equation}
\label{slowly-varying}
\left(\frac{\dd}{\dd z} + \frac{1}{v_g}
\frac{\dd}{\dd t}
\right){\cal E}(z,t) = i {k \over 2} \chi(\nu){\cal E}(z,t)~.
\end{equation}

	Traditionally, the susceptibility is divided into
real and imaginary parts: $\chi = \chi' + i \chi''$, and
vanishingly small $\chi'$ and $\chi''$ at resonance are the
signature of EIT.  The group velocity in the medium is given by
$
v_g = {c/[1 + (\nu/2)(d\chi'/d\nu)]}~,
$
where the derivative is evaluated at the carrier frequency.

	For a medium displaying EIT, $\chi(\nu_p)$ is given
by~\cite{scullybook}
\begin{equation}
 \label{t1}
 \chi(\nu_p) =
 	\int_{-\infty}^{\infty}
	   {  i\,\eta\,\gamma_{r}\Gamma_{bc}
	   	\over
              \Gamma_{bc} [\gamma + i (\Delta_p + k_p v) ] + \Omega^2
	   } \, f(v) dv~.
\end{equation}
In this expression $\nu_p$ is the probe laser frequency,
$\eta = (3\lambda^3 N)/(8 \pi^2)$ where $\lambda$ is the
probe wavelength and $N$ is the atomic density, $\Gamma_{bc}
= \gamma_{bc} + i[\delta + (k_p - k_d)v]$ where $\gamma_r$
is the radiative decay rate of level $a$ to level $b$,
$\gamma_{bc}$ is the coherence decay rate of the two lower
levels, (governed here by the time-of-flight through the laser
beams), $\gamma$ is the total homogeneous half-width of the
drive and probe transitions (including radiative decay and
collisions), $\Delta_p = \omega_{ab} - \nu_p$ and $\Delta_d =
\omega_{ac} - \nu_d$ are the one-photon detunings of the probe
and drive lasers, and $\delta = \Delta_p - \Delta_d$ is the
two-photon detuning, $\Omega$ is the Rabi frequency of the drive
transition, $k_p$ and $k_d$ are wave numbers of the probe and
driving fields respectively.  One can obtain a simple analytic
expression corresponding to Eq.~(\ref{t1}) by approximating
the thermal distribution $f(v)$ by a lorentzian, $f(v) =
(kv/\pi)/[(\Delta\omega_D)^2 + (kv)^2]$, where $\Delta\omega_D$
is the Doppler half-width of the thermal distribution and $v$
is the projection of the atomic velocity along the laser beams.
The result is
\begin{equation}
 \label{lorentz-avg}
 \chi(\nu_p) = \eta\, \gamma_{r}
	     \frac{i\gamma_{bc} - \delta}
	          {(\gamma + \Delta\omega_D + i\Delta_p)
		   (\gamma_{bc} + i\delta)
		   +\Omega^2
		  }
\end{equation}
where we have taken $k = k_p = k_d$.

Equation~(\ref{lorentz-avg}) leads to propagation with
absorption coefficient $\alpha = (k/2)\chi''(\nu_p)$ and group
velocity $v_g = c/(1+n_g)$.  We obtain
\begin{eqnarray}
\alpha &=& \frac{3}{8\pi} \, N \lambda^2 \,
		\frac{\gamma_r \gamma_{bc} }
		     {\gamma_{bc}(\gamma + \Delta\omega_D) + \Omega^2}, \\
\label{ng-on-omega}
n_g &=& \frac{3}{8\pi} \, N \lambda^2 \,
		\frac{\gamma_r \Omega^2 c}
		     {[\gamma_{bc}(\gamma + \Delta\omega_D) + \Omega^2]^2}~.
\end{eqnarray}
After propagation through a dense coherent ensemble of
length $L$ the intensity of the pulse is attenuated by
$\exp(-2\alpha L)$, whereas its envelope is delayed compared
to free space propagation by $T_g = n_g L/c\;$.

	To relate the present results to earlier studies of
EIT-based spectroscopy and prior group delay measurements,
we note that the group delay is essentially the reciprocal
of the so-called ``dispersive width'' associated with the EIT
resonance $\Delta \omega_{dis} = \pi/(2 T_g)$.  This width was
defined in Ref.\cite{lukin97prl} as the detuning from the line
center at which the phase of the probe laser shifts by $\pi/2$.
The significance of this quantity is that it determines the
ultimate resolution of interferometric measurements using
EIT.  When the group delay is large, the dispersive width is
correspondingly small.  This is the basis for high-precision
spectroscopy in dense coherent media.

	Clearly, an essential difference between hot
and cold atom experiments concerns Doppler broadening.
Equation~(\ref{ng-on-omega}) shows that for our experimental
regime where $\Omega^2 \gg \gamma_{bc}(\gamma + \Delta\omega_D)$
the effect of Doppler averaging is not important.  The point
is that in many current experiments (EIT, LWI, high resolution
dense medium spectroscopy, and ultra-slow group velocities)
the results are two-photon Doppler free for copropagating
drive and probe fields.  That is, as shown in Eq.~(\ref{t1}),
when $k_p \approx k_d$, only the single photon denominator
(the square-bracketed expression in Eq.~(\ref{t1})) depends
on atomic velocity.  This has a negligible effect near two
photon resonance, provided that the Rabi-frequency of the
driving field is sufficiently large.  This analysis, our
experimental demonstration, and numerical calculations, allow
us to conclude that for strong driving fields, the effects
of Doppler averaging are not of central importance to the
group velocity.  The results of numerical calculations are
shown in Fig.~\ref{dopp.fig}.  For our current experiments,
in which $\gamma_{bc}/(2\pi) \approx 10^{3}$ Hz and $(\gamma
+ \Delta\omega_D)/(2\pi) \approx 4\times10^{8}$ Hz, drive
Rabi frequencies $\Omega/(2\pi) \gg 10^{6}$ Hz are required
in order to get a measurable signal through the cell.
We note from Fig.~\ref{dopp.fig} that for this range of
intensity, the lorentzian approximation holds, and also that
$\partial\chi'/\partial\nu$ is nearly the same for hot and
cold gases.  Furthermore, curve (d) in Fig.~\ref{dopp.fig}
shows that for our current experimental conditions, a reduction
of $\gamma_{bc}$ allows the possibility of reaching much
lower group velocities, near 10 m/s.  Such a reduction in
$\gamma_{bc}$ is quite possible by increasing our laser beam
diameter as shown in Ref.~\cite{wynands-narrow}.

	On the other hand, the cold atom technology
does hold promise for a truly Doppler free payoff; e.g.,
nonlinear optical processes involving ``sideways'' coupling
\cite{hau99nature,Harris-NLO-weak} in which the drive and
probe lasers are perpendicular.  This is not possible in a
hot gas.  Likewise, EIT experiments in cold gases might be a
very interesting tool for studying the properties of and even
manipulating the Bose condensate.

\begin{figure}[ht]
 \centerline{\epsfig{file=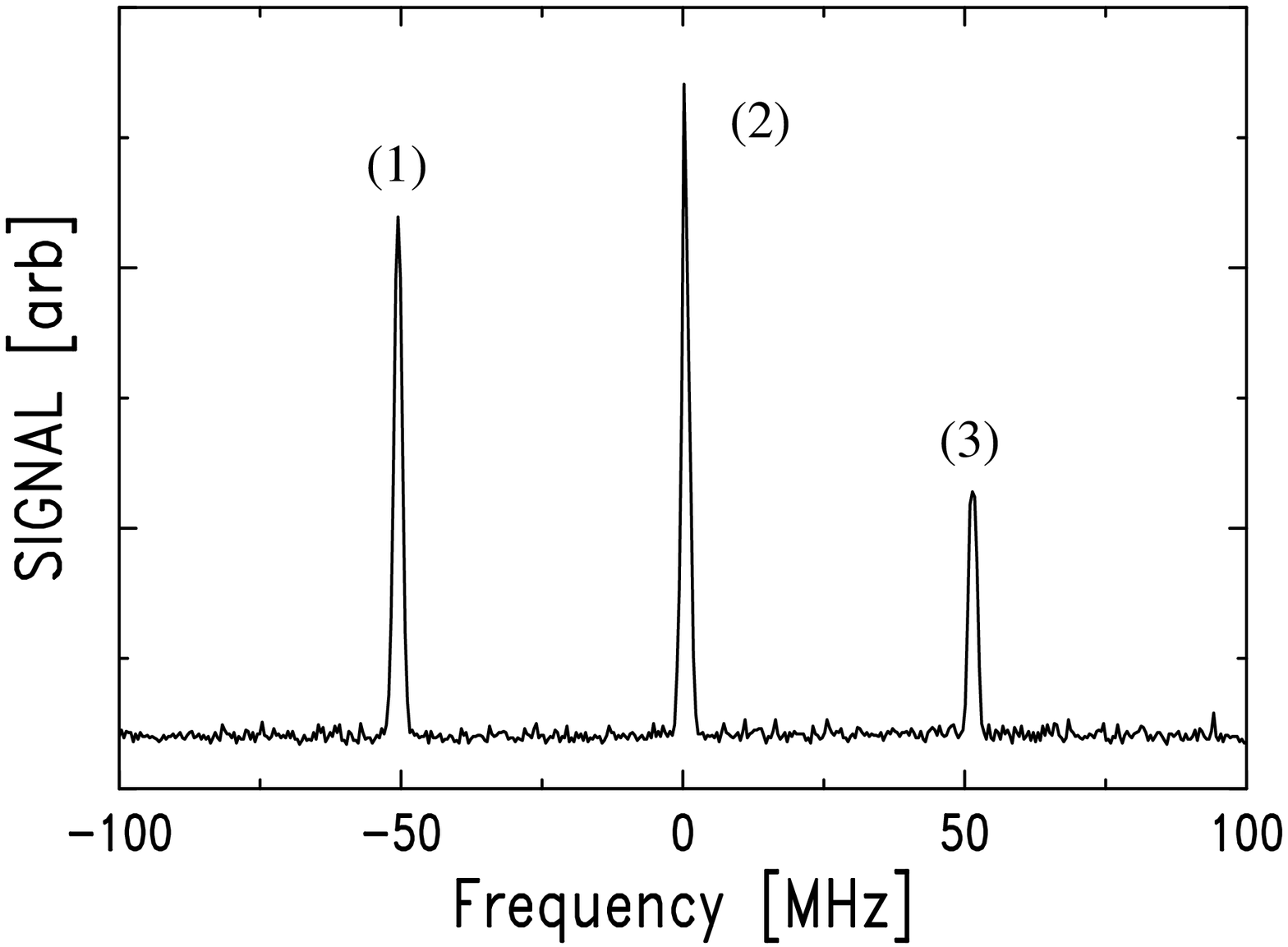,width=7.0cm}}
 \vspace*{2ex}
 \caption{
 	\label{new-field-data.fig}
	RF spectrum for frequencies in the vicinity of the
	ground-state hyperfine splitting showin build-up of a
	``new''-field from the vacuum, plotted on a linear
	scale.	Peak (1) is the beat between the probe and
	shifted fields, peak (2) is the beat between the
	drive, probe, and new fields, and peak (3) is the
	beat between the new and shifted fields.  Note that
	the amplitude of the the new field is comparable to
	the probe field indicating {\it extremely} efficient
	nonlinear generation.  The drive power was 4.3 mW.
}
\end{figure}

	We next turn to nonlinear interactions such as wave
mixing involving pulses or cw fields in a phase coherent media.
In the present system, nonlinear wave mixing phenomenon can
be induced by the off-resonant coupling of states $a$ and $b$
in Fig.~\ref{levs-expt.fig}a by the driving field, or by
applying a second driving field $\Omega_2$.  As discussed
previously~\cite{lukin97prl,lukin98prl} two important
situations should be distinguished.  They correspond to driving
fields with Rabi frequencies $\Omega_1$ and $\Omega_2$ that
propagate in the same or opposite directions, respectively.
In the case of counter-propagating fields, oscillation
occurs~\cite{zib-osc-submitted}.  In the case of co-propagating
fields at high optical power and Rb vapor density, coherent
Raman scattering leads to efficient nonlinear generation of a
Stokes component (peak (3) in Fig.~\ref{new-field-data.fig}).
The importance of Raman nonlinearities in such resonant wave
mixing phenomena has been discussed previously~\cite{Boyd}.

	Evidence of this generation for the case
of co-propagating fields has been observed for cw
fields~\cite{lukin97prl}.  There, the off-resonant coupling of
the drive field to the $a\leftrightarrow b$ transition led to
generation of a new (Stokes) field.  The new field is generated
at $\nu_d - \omega_{cb}$ (see Fig.~\ref{levs-expt.fig}a).
In Ref.~\cite{lukin97prl} this new field was observed indirectly
by observing the simultaneous beat between the drive, probe,
and new fields at $\omega_{cb}$.

	By using the method described above for isolating
the strength of the transmitted probe field, we can
now isolate the strength of the generated new field.
Figure~\ref{new-field-data.fig} shows the relative strengths
of the probe and new field.  It is striking to note that,
under conditions of a large group delay, the output power in
these two fields (1 and 3) is nearly the same.  Furthermore,
the properties of this field may be studied with the current
experimental arrangement with a time-varying probe, and will
be discussed in detail elsewhere.

	We emphasize the connection between ultra-slow
light propagation and large nonlinearities.  As in
Fig.~\ref{new-field-data.fig}, consider the case of co-linear
propagation of the slowly varying anti-Stokes (i.e. probe) and
Stokes (i.e. new) fields, coupled by coherent Raman scattering.
We assume that the cw driving field is near resonance with the
$c\leftrightarrow a$ transition.  In the vicinity of two-photon
resonance the probe field ${\cal E}_p$ and the new Stokes field
${\cal E}_n$ evaluated at the exit from the cell of length $L$
is given by \cite{lukin98prl}:
\begin{eqnarray}
{\cal E}_p   &=&  {\cal E}_p(0) {\rm cosh}(\xi T_g) \\
{\cal E}_n^* &=& i{\cal E}_p(0) {\rm sinh}(\xi T_g)
\label{why_this_one}
\end{eqnarray}
where $\xi = \Omega^2/\omega_{cb}$ and for simplicity we
have ignored loss and chosen the detuning $\delta$ such
that phase-matching is satisfied (i.e.\ $\delta = (k_p +
k_n - 2k_d)c/n_g)$.  This indicates that the medium with a
sufficiently long-lived ground state coherence $\gamma_{bc}$,
and sufficiently large density-length product (to insure large
group delay, and consequently large nonlinear gain) is required
to achieve efficient nonlinear generation. We emphasize that
these requirements are typical for any efficient nonlinear
interactions involving phase coherent media.  Dramatic examples
are large Kerr nonlinearities~\cite{atac98oc}, single photon
switching \cite{harris-yam-98-prl}, and quantum control and
correlations of weak laser beams~\cite{nonlinear-lowlight}.

	In conclusion we have demonstrated ultra-large
group delay $T_g \approx 0.26$ ms for light traversing a
cell containing an ensemble of hot phase coherent atoms for
which the transit time through an empty cell is a fraction of
a nanosecond.  Such a phaseonium gas has ultra-large nonlinear
optical properties yielding nonlinear coupling between very
weak fields.  It is safe to predict that such phase coherent
materials will be of both fundamental (e.g.\ probing the Bose
condensate) and applied (e.g.\ compression of information by
many orders of magnitude) interest.

	This work was supported by the Office of Naval Research,
the National Science Foundation, the Robert A. Welch Foundation,
and the U.\ S.\ Air Force.  We thank Steve Harris for his
enthusiasm concerning the present studies resulting in this
paper, and G.\ Agarwal, C.\ Bednar, S.\ Harris, L.\ Hau, H.\
Lee, and A.\ Matsko for stimulating and helpful discussions.

\def\etal{\textit{et al.}}


\end{document}